\pdfoutput=1
\documentclass[preprint,showpacs,preprintnumbers,amsmath,amssymb,floatfix,endfloats*]{revtex4}

\usepackage{graphicx}
\usepackage{dcolumn}
\usepackage{bm}
\usepackage{amsfonts}

\DeclareMathAlphabet{\mathsfsl}{OT1}{cmr}{bx}{it}
\begin{document}
\title{Dynamical heterogeneity in periodically deformed polymer glasses}
\author{Nikolai V. Priezjev}
\affiliation{Department of Mechanical and Materials Engineering,
Wright State University, Dayton, OH 45435}
\date{\today}
\begin{abstract}

The dynamics of structural relaxation in a model polymer glass
subject to spatially-homogeneous, time-periodic shear deformation is
investigated using molecular dynamics simulations.    We study a
coarse-grained bead-spring model of short polymer chains   below the
glass transition temperature.   It is found that at small strain
amplitudes, the segmental dynamics is nearly reversible over about
$10^4$ cycles, while at strain amplitudes above a few percent,
polymer chains become fully relaxed after a hundred cycles.    At
the critical strain amplitude, the transition from slow to fast
relaxation dynamics is associated with the largest number of
dynamically correlated monomers as indicated by the peak value of
the dynamical susceptibility.   The analysis of individual monomer
trajectories showed that mobile monomers tend to assist their
neighbors to become mobile and aggregate into relatively compact
transient clusters.

\end{abstract}



\maketitle

\section{Introduction}

The analysis and optimization of the mechanical performance of
amorphous polymers are critical for various industrial and
biomedical applications~\cite{Meijer05}.   In the absence of
external deformation, the molecular motion slows down and a polymer
glass gradually evolves towards an equilibrium state in a process
called {\it physical aging}, which affects mechanical properties of
the material~\cite{Struik78}.    In turn, the effects of physical
aging can be removed by application of mechanical stresses or by
heating above the glass transition temperature and then cooling back
down~\cite{Struik78}.    According to the well-known Eyring model,
an applied stress lowers the effective energy barriers for molecular
motion and thus induces yield and plastic flow in polymer
glasses~\cite{Eyring36}.   This simple description, however, does
not include the effects of dynamical heterogeneity, strain
localization, and strain hardening~\cite{BarratCh8,Hoy06}.

It was previously demonstrated that the relaxation dynamics in {\it
quiescent} polymer glasses near the glass transition temperature
becomes spatially
heterogeneous~\cite{GlotNat99,GlotStar01,Aichele03}.    In
particular, it was shown that the most mobile monomers form
transient clusters whose mean size increases upon cooling towards
the glass transition temperature~\cite{GlotStar01}.    In some
cases, neighboring monomers undergo large displacements and follow
each other in a string-like fashion~\cite{Aichele03}.   Near $T_g$,
the average string length was found to be about two monomer
diameters, although strings of about ten monomers were
observed~\cite{Aichele03}.    As expected, polymer chain ends are
more mobile; however, the mobility does not necessarily propagate
along the backbone of the chains~\cite{Aichele03}.  More recently,
the spatiotemporal distribution of monomer hopping events was
investigated in an aging polymer glass quenched below the glass
transition temperature~\cite{Rottler13}.    It was shown that before
merging into a single dominating cluster, the volume distribution of
clusters of hopping monomers follows a power-law decay with an
exponent of two, and the clusters have slightly noncompact
shapes~\cite{Rottler13}.

The segmental mobility during {\it constant stress or strain rate
deformation} of polymer glasses was recently studied
experimentally~\cite{Gleason00,Ediger08,Ediger09} and using
molecular dynamics (MD)
simulations~\cite{Rutledge02,Lyulin05,Papa08,dePablo10,RottlerJCP10,RottlerPRL10}.
In general, it was shown that after flow onset, the segmental
mobility is strongly accelerated and the distribution of relaxation
times is narrowed under active deformation.   It was also found that
before the onset of flow, the deformation-induced molecular mobility
is spatially heterogeneous, involving the formation of clusters of
mobile molecules~\cite{dePablo10}. By decomposing monomer
trajectories into a series of rapid hopping events, it was observed
that the distribution of the first hop and persistence times is
narrowed, which indicates that the monomer relaxation dynamics is
accelerated during constant strain rate
deformation~\cite{RottlerJCP10,RottlerPRL10}. As a complimentary
approach to probe glassy dynamics, the analysis of individual
particle trajectories was also performed in amorphous materials
under {\it cyclic shear
deformation}~\cite{BiroliPRL09,Sastry13,Priezjev13,OHern13,Lyulin13,Reichhardt13,Arratia13}.
By employing a novel cage decomposition algorithm, it was
demonstrated that intermittent bursts of cage jumps are directly
correlated with the major structural relaxation events in a
two-dimensional dense granular media~\cite{BiroliPRL09}.      It was
further pointed out that the relaxation process involves spatially
clustered cage jumps, which on long time scales aggregate into
avalanches~\cite{BiroliPRL09}.

In a recent study~\cite{Priezjev13}, the relaxation dynamics in
periodically deformed binary Lennard-Jones mixture was examined at a
finite temperature well below the glass transition.  It was found
that the mean-square displacement develops an extended subdiffusive
plateau associated with cage-trapping, and the particle dynamics
becomes spatially and temporally heterogeneous.     Furthermore, the
dynamic correlation length, which was estimated from the peak of the
dynamical susceptibility, grows with increasing strain amplitude up
to a value that corresponds to the largest size of dynamically
correlated regions.    One of the aims of the current study is to
test whether these conclusions hold for nonentangled polymer glasses
under cyclic loading.

In this paper, we investigate structural relaxation and dynamical
heterogeneity in a bead-spring model of low-molecular-weight polymer
glass that is subject to spatially homogeneous, time-periodic shear
deformation.     We find that at sufficiently small strain
amplitudes, the system dynamics is nearly reversible, while at
amplitudes above a few percent, monomers undergo irreversible cage
jumps that become spatially aggregated into relatively compact
clusters.    It will be shown that at the critical strain amplitude,
the dynamic correlation length exhibits a distinct maximum
indicating the largest size of regions over which the motion of
monomers is spatially correlated.

The rest of the paper is structured as follows.  The description of
molecular dynamics simulations is presented in
Sec.\,\ref{sec:MD_Model}.  In Sec.\,\ref{sec:Results}, we examine
the mean-square displacement of monomers, the autocorrelation
function of normal modes, as well as the self-correlation function
and dynamical susceptibility, followed by the analysis of the
monomer hopping dynamics and the discussion of dynamic facilitation.
Brief conclusions are provided in the final section.

\section{Details of molecular dynamics simulations}
\label{sec:MD_Model}

\begin{figure}[t]
\includegraphics[width=12.cm,angle=0]{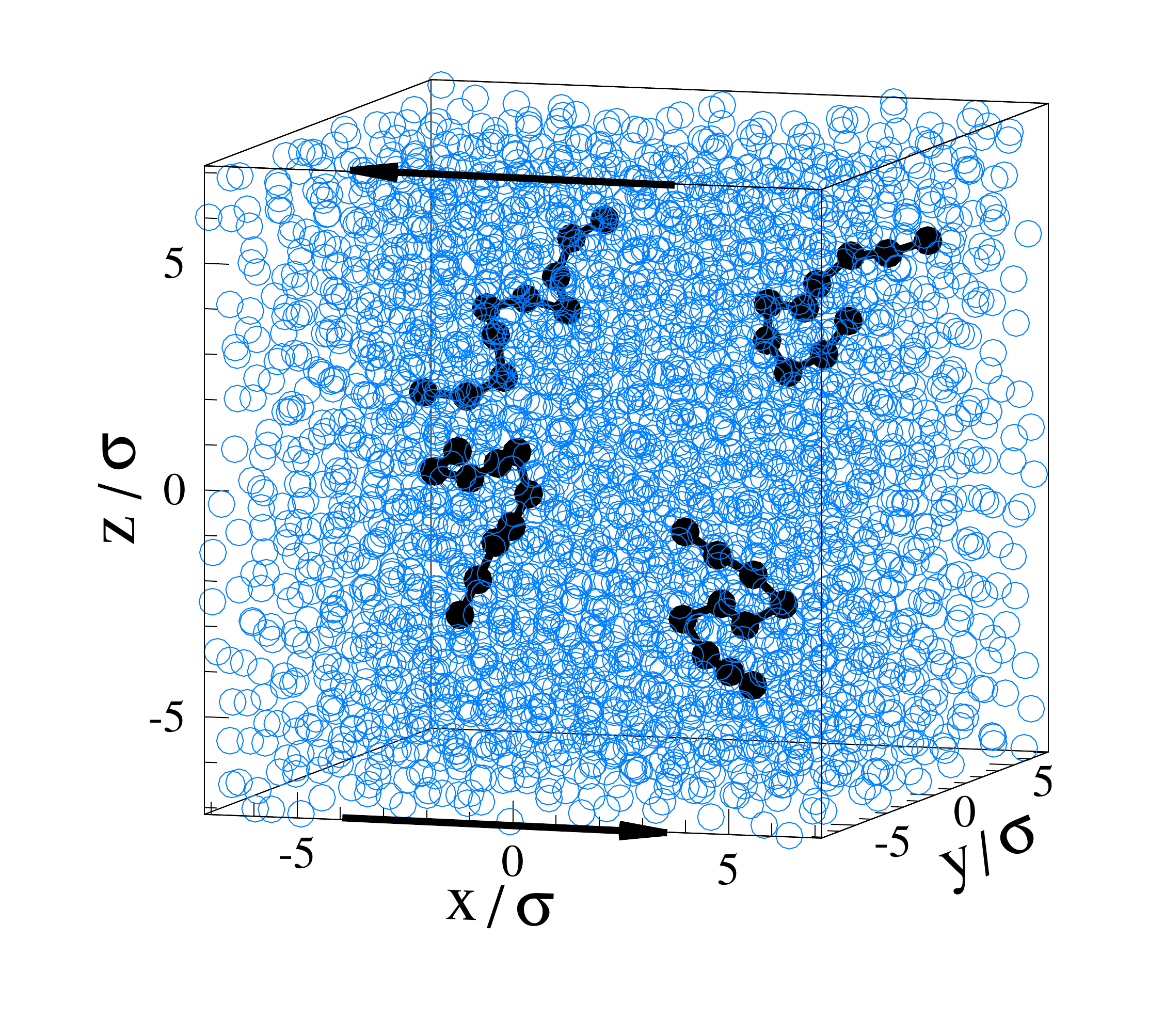}
\caption{(Color online)   A snapshot of the polymer glass during
homogeneous time-periodic shear deformation with the strain
amplitude $\gamma_{0}=0.02$ and frequency $\omega\tau=0.05$.  Four
chains of $N=10$ monomers are marked by solid lines and filled
circles (not drawn to scale).   The black arrows indicate the
direction of the applied shear strain.   The Lees-Edwards periodic
boundary conditions are imposed in the $xz$ plane. }
\label{fig:snapshot_system}
\end{figure}

In this study, we perform molecular dynamics simulations of the
coarse-grained bead-spring model of flexible polymer
chains~\cite{Kremer90}.   The system consists of $312$ linear chains
of $N=10$ monomers each confined in a periodic cubic cell.   A
snapshot of the polymer glass at zero strain is shown in
Fig.\,\ref{fig:snapshot_system}.   The pairwise interaction between
monomers is specified by the truncated Lennard-Jones (LJ) potential
\begin{equation}
V_{LJ}(r)=4\,\varepsilon\,\Big[\Big(\frac{\sigma}{r}\Big)^{12}\!-\Big(\frac{\sigma}{r}\Big)^{6}\,\Big],
\label{Eq:LJ}
\end{equation}
where the parameters $\varepsilon$ and $\sigma$ denote the energy
and length scales, respectively.    The cutoff radius is fixed to
$r_{c}=2.245\,\sigma$.  The total number of monomers in the system
is $N_m=3120$.   In addition to the LJ potential, any two
neighboring beads in a polymer chain interact via the finitely
extensible nonlinear elastic (FENE) potential~\cite{Bird87}
\begin{equation}
V_{FENE}(r)=-\frac{k_s}{2}\,r_{\!o}^2\ln[1-r^2/r_{\!o}^2],
\label{Eq:FENE}
\end{equation}
with the parameters $k_s=30\,\varepsilon\sigma^{-2}$ and
$r_{\!o}=1.5\,\sigma$~\cite{Kremer90}.   The effective bond
potential between neighboring beads does not allow chain crossings
and bond breaking even at the highest strain amplitude considered in
the present study.

The simulations were carried out at a constant temperature of
$0.1\,\varepsilon/k_B$, which is below the glass transition
temperature $T_g\approx0.32\,\varepsilon/k_B$~\cite{Binder98}. Here,
$k_B$ is the Boltzmann constant.   To keep the system temperature at
$0.1\,\varepsilon/k_B$, the velocity component perpendicular to the
plane of deformation was rescaled every $10$ MD steps.   The polymer
glass was confined into a cubic box with a side length of
$14.29\,\sigma$, resulting in a monomer density
$\rho=1.07\,\sigma^{-3}$ (see Fig.\,\ref{fig:snapshot_system}).
A homogeneous shear deformation was imposed using the SLLOD
algorithm combined with the Lees-Edwards periodic boundary
conditions~\cite{Evans92}.    The equations of motion were
integrated using the fifth-order Gear predictor-corrector
algorithm~\cite{Allen87} with a time step $\triangle
t_{MD}=0.005\,\tau$, where $\tau=\sigma\sqrt{m/\varepsilon}$ is the
LJ time.


After equilibration for about $5\times10^6$ MD steps, the cyclic
shear strain was applied in the $xz$ plane by varying strain as a
function of time as follows
\begin{equation}
\gamma(t)=\gamma_{0}\,\,\textrm{sin}(\omega t), \label{Eq:strain}
\end{equation}
where $\gamma_{0}$ is the strain amplitude and $\omega$ is the
oscillation frequency.    In our simulations, the oscillation
frequency and period were fixed to $\omega\tau=0.05$ and
$T=2\pi/\omega=125.66\,\tau$, respectively.    The maximum strain
amplitude considered in the present study, $\gamma_{0} = 0.09$, is
greater than the yield strain.   After discarding transients, the
positions of all monomers were saved every back and forth cycle when
the net strain is zero, and the data were gathered over $15\,000$
cycles (about $3.8\times10^8$ MD steps) at each strain amplitude.
Therefore, as the cyclic loading continues, the structural changes
in the material are related to the degree of overlap between monomer
configurations at different times. The post-processing analysis of
the MD data was carried out in ten independent samples for each
$\gamma_{0}$.

\section{Results}
\label{sec:Results}

\begin{figure}[t]
\includegraphics[width=12.cm,angle=0]{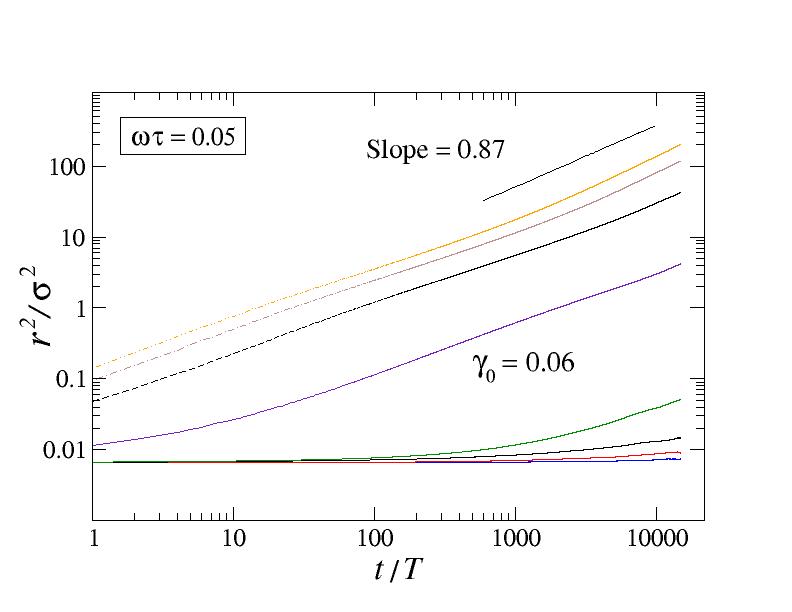}
\caption{(Color online)   The averaged mean-square displacement of
monomers as a function of time for the oscillation frequency
$\omega\tau=0.05$ and period $T=2\pi/\omega=125.66\,\tau$.  The
strain amplitudes from bottom to top are
$\gamma_{0}=0.02,~0.03,~0.04,~0.05,~0.06,~0.07,~0.08,~\text{and}~0.09$.
The straight black line with the slope $0.87$ is shown for
reference. } \label{fig:msd_time_omega}
\end{figure}

The molecular structure of amorphous polymers is characterized by
the short-range order and the absence of any long-range order or
symmetry~\cite{Paul04}.    During the time-periodic, steady-state
deformation, monomers of a polymer chain either remain trapped
within cages formed by their neighbors or undergo irreversible
displacements, which gives rise to a diffusion process.
Figure\,\ref{fig:msd_time_omega} shows the time dependence of the
mean-square displacement of monomers for different strain
amplitudes.   Before averaging, the displacement vector for each
monomer was computed with respect to the center of mass of the whole
system.  It can be seen that with increasing strain amplitude, the
characteristic time for the onset of the diffusive motion decreases.
Notice that at small strain amplitudes,
$\gamma_{0}=0.02~\text{and}~0.03$, monomers remain trapped in their
cages during the time interval $15\,000\,T$, while at larger
amplitudes, $\gamma_{0}=0.04~\text{and}~0.05$, monomers escape from
their cages after about $1000\,T$.    When
$\gamma_{0}\geqslant0.06$, the monomer dynamics is slightly
subdiffusive at long times as the displacement of monomers is
restricted by the motion of the center of mass of polymer chains.
Interestingly, at large strain amplitudes $\gamma_{0}\geqslant0.07$,
monomers, on average, move out from their cages during a single
oscillation cycle.   Lastly, the ballistic regime occurs at shorter
time scales than the oscillation period $T=125.66\,\tau$, and,
therefore, it is not present in any of the curves in
Fig.\,\ref{fig:msd_time_omega}.

\begin{figure}[t]
\includegraphics[width=12.cm,angle=0]{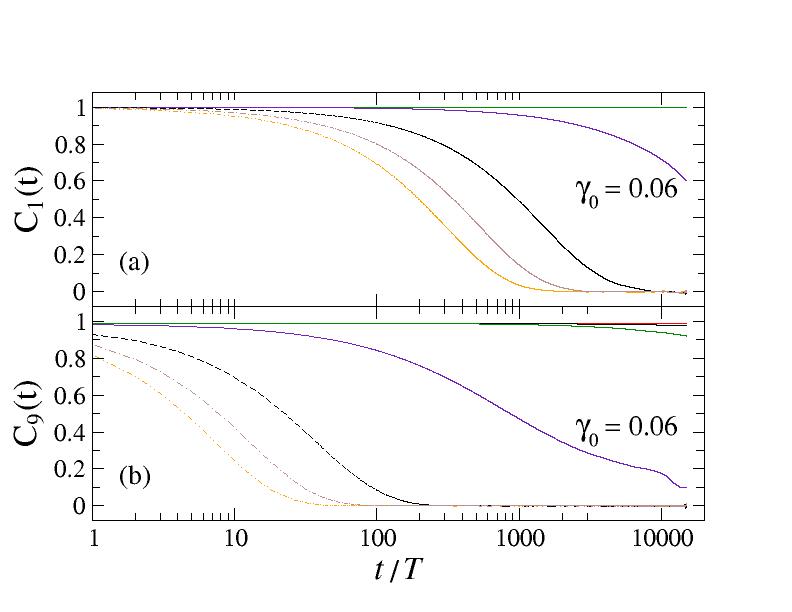}
\caption{(Color online) The autocorrelation function of (a) $p=1$
and (b) $p=9$ normal modes defined by Eq.\,(\ref{Eq:auto_corr_eq})
for the oscillation frequency $\omega\tau=0.05$ and period
$T=2\pi/\omega=125.66\,\tau$.   The strain amplitudes from top to
bottom are
$\gamma_{0}=0.02,~0.03,~0.04,~0.05,~0.06,~0.07,~0.08,~\text{and}~0.09$.}
\label{fig:norm_modes}
\end{figure}

The relaxation dynamics of polymer chains can be probed by analyzing
the decay of the time autocorrelation function of normal
modes~\cite{Verdier66,Varnik05}.  For a polymer chain that consists
of $N$ monomers, the normal coordinates are defined by
\begin{equation}
\mathbf{X}_p(t)=\frac{1}{N}\sum_{i=1}^N
\mathbf{r}_i(t)\,\textrm{cos}\,\frac{p\pi(i-1/2)}{N},
\label{Eq:normal_mode_eq}
\end{equation}
where $\mathbf{r}_i$ is the position vector of the $i$-th monomer
and $p=0, 1,..., N-1$ is the mode number.  The shortest and longest
relaxation times correspond to the last $p=N-1$ and the first $p=1$
modes.    Correspondingly, the time autocorrelation function for the
$p$-th normal mode is computed as follows:
\begin{equation}
C_p(t)=\langle \mathbf{X}_p(t)\cdot \mathbf{X}_p(0)\rangle/ \langle
\mathbf{X}_p(0)\cdot \mathbf{X}_p(0)\rangle, \label{Eq:auto_corr_eq}
\end{equation}
where the angle brackets denote averaging over initial times and
independent samples.    The time dependence of the correlation
functions $C_1(t)$ and $C_9(t)$ is presented in
Fig.\,\ref{fig:norm_modes} for the oscillation frequency
$\omega\tau=0.05$ and different strain amplitudes.   The
orientational dynamics of the whole chain is described by the
function $C_1(t)$.   It is evident from
Fig.\,\ref{fig:norm_modes}\,(a), that the orientation of polymer
chains is unaffected by the periodic deformation for strain
amplitudes $\gamma_{0}\leqslant0.05$, while they become fully
relaxed for $\gamma_{0}\geqslant0.07$ during the reported time
interval.   As expected, the segmental dynamics is faster; e.g., the
function $C_9(t)$ decays to nearly zero after $1.5\times10^4$ cycles
for the strain amplitude $\gamma_{0}=0.06$, as shown in
Fig.\,\ref{fig:norm_modes}\,(b).   These results indicate that with
increasing amplitude of the shear strain deformation, the relaxation
dynamics of polymer chains undergoes a transition at the stain
amplitude of about $\gamma_{0}\approx0.06$.

\begin{figure}[t]
\includegraphics[width=12.cm,angle=0]{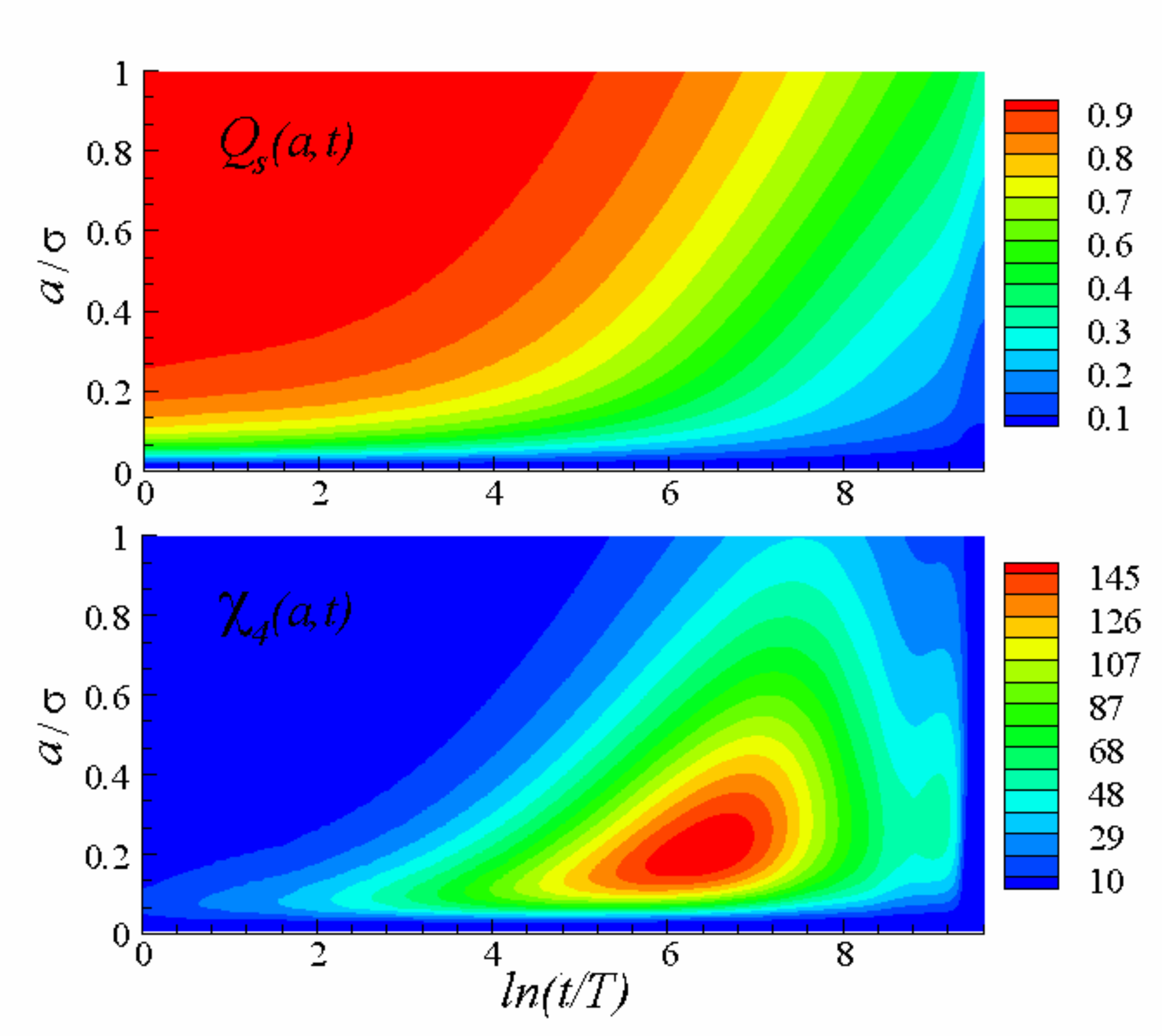}
\caption{(Color online)  The contour plots of the correlation
functions $Q_s(a,t)$ (top) and $\chi_4(a,t)$ (bottom) for the strain
amplitude $\gamma_{0}=0.06$ and oscillation frequency
$\omega\tau=0.05$.  The oscillation period is $T=125.66\,\tau$. }
\label{fig:q_k_map}
\end{figure}

The structural relaxation process in glassy materials often involves
spatial fluctuations of particle mobilities~\cite{Berthier11}.
During periodic deformation, the degree of overlap between two
spatial configurations of monomers is described by the
self-correlation function, which is defined as follows:
\begin{equation}
Q_s(a,t)=\frac{1}{N_{m}} \sum_{i=1}^{N_{m}}\text{exp}\Big( -
\frac{\Delta\mathbf{r}_{i}(t)^2}{2\,a^{2}} \Big),
\label{Eq:self_corr}
\end{equation}
where
$\Delta\mathbf{r}_{i}(t)=\mathbf{r}_{i}(t_0+t)-\mathbf{r}_{i}(t_0)$
is the displacement vector of the $i$-th monomer during the time
interval $t$, $N_m$ is the total number of monomers, and $a$ is the
probed length scale~\cite{BiroliPRL05}.   Furthermore, it was
previously shown that the dynamical heterogeneity can be quantified
via the variance of the self-correlation function, or the dynamical
susceptibility, which is given by
\begin{equation}
\chi_4 (a,t) = N_{m}\,\big[ \langle Q_s(a,t)^2 \rangle - \langle
Q_s(a,t) \rangle^2 \big],    \label{Eq:four_point_corr}
\end{equation}
where the averaging is performed over all initial
times~\cite{Glotzer00}.   An example of the correlation functions
$Q_s(a,t)$ and $\chi_4 (a,t)$ is presented in
Fig.\,\ref{fig:q_k_map} for the strain amplitude $\gamma_{0}=0.06$.
The contour plots clearly show that during the reported time
interval the structural relaxation occurs on the length scale of
about the cage size, and the dynamical susceptibility $\chi_4 (a,t)$
reaches a maximum at intermediate length and time scales, thus
providing an estimate for a number of monomers involved in a
correlated motion.

\begin{figure}[t]
\includegraphics[width=12.cm,angle=0]{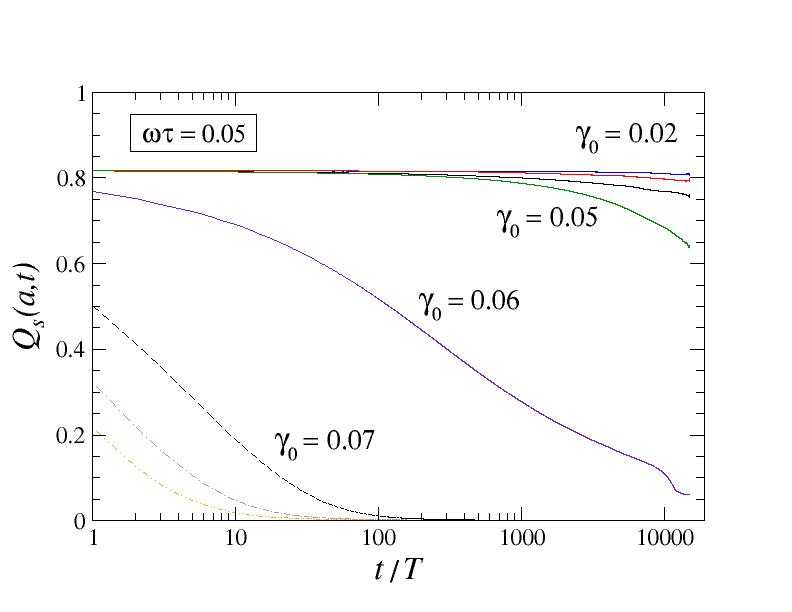}
\caption{(Color online)    The time dependence of the
self-correlation function $Q_s(a,t)$ computed at $a=0.12\,\sigma$
for the oscillation frequency $\omega\tau=0.05$ and period
$T=2\pi/\omega=125.66\,\tau$.    The strain amplitudes from top to
bottom are
$\gamma_{0}=0.02,~0.03,~0.04,~0.05,~0.06,~0.07,~0.08,~\text{and}~0.09$.}
\label{fig:q28_time_omega_0.05}
\end{figure}

To gain further insight into the relaxation process, we fix the
probed length scale to a value slightly larger than the cage size,
i.e., $a=0.12\,\sigma$, and plot the self-correlation function
$Q_s(a,t)$ versus time in Fig.\,\ref{fig:q28_time_omega_0.05} for
different strain amplitudes.   It is clearly observed that the
structural relaxation occurs faster at larger strain amplitudes.
In particular, at small strain amplitudes,
$\gamma_{0}\leqslant0.02$, the monomers remain trapped inside their
cages, indicating that the system dynamics is nearly reversible
during the reported time interval; while for
$\gamma_{0}\geqslant0.07$, the system becomes fully relaxed after
about $100$ cycles.    Similar to the behavior of the
autocorrelation function of normal modes shown in
Fig.\,\ref{fig:norm_modes}, the transition from slow to fast
dynamics occurs at the same strain amplitude
$\gamma_{0}\approx0.06$.   Also, the results in
Fig.\,\ref{fig:q28_time_omega_0.05} are consistent with the time
dependence of the mean-square displacement curves reported in
Fig.\,\ref{fig:msd_time_omega}.

\begin{figure}[t]
\includegraphics[width=12.cm,angle=0]{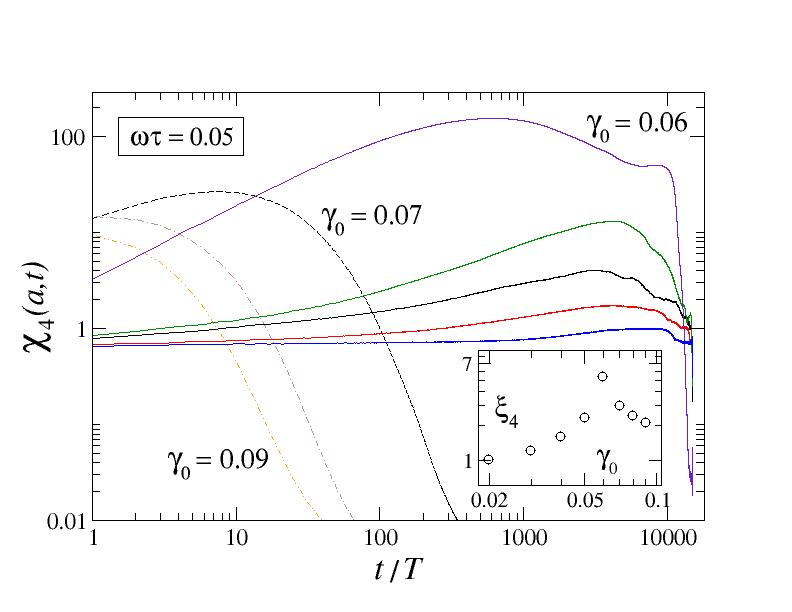}
\caption{(Color online)      The dynamic susceptibility
$\chi_4(a,t)$ defined by Eq.\,(\ref{Eq:four_point_corr}) for the
oscillation frequency $\omega\tau=0.05$ and period
$T=2\pi/\omega=125.66\,\tau$.     The strain amplitudes from bottom
to top are $\gamma_{0}=0.02~\text{at}~a=0.06\,\sigma$,
$\gamma_{0}=0.03~\text{at}~a=0.06\,\sigma$,
$\gamma_{0}=0.04~\text{at}~a=0.07\,\sigma$,
$\gamma_{0}=0.05~\text{at}~a=0.08\,\sigma$,
$\gamma_{0}=0.06~\text{at}~a=0.21\,\sigma$.  The other curves
correspond to $\gamma_{0}=0.07~\text{at}~a=0.20\,\sigma$ (dashed
curve), $\gamma_{0}=0.08~\text{at}~a=0.17\,\sigma$ (dash-dotted
curve), and $\gamma_{0}=0.09~\text{at}~a=0.19\,\sigma$
(dash-double-dotted curve).     The inset shows the dynamic
correlation length $\xi_{4}=[\chi_4^{\text{max}}(a,t)]^{1/3}$ as a
function of the strain amplitude $\gamma_{0}$.  }
\label{fig:ksi28_time_omega_0.05}
\end{figure}

While analyzing the dynamical susceptibility at different strain
amplitudes, we found that for each $\gamma_{0}$ the location of the
maximum of $\chi_4(a,t)$ depends both on $a$ and $t$.    In
Fig.\,\ref{fig:ksi28_time_omega_0.05}, the dynamical susceptibility
is shown as function of time for the values of the parameter $a$ at
which $\chi_4(a,t)$ has a global maximum at a given strain
amplitude.   It is apparent that with increasing strain amplitude,
the position of the peak in $\chi_4(a,t)$ is shifted to smaller
times.   The amplitude of the peak, which reflects the typical
number of monomers involved in a correlated motion, has a pronounced
maximum at the strain amplitude $\gamma_{0}=0.06$.  Notice also that
at larger strain amplitudes, $\gamma_{0}=0.08~\text{and}~0.09$, the
maximum of $\chi_4(a,t)$ occurs after the first cycle.


Assuming that correlated regions are relatively compact, the dynamic
correlation length $\xi_4$ can be simply estimated from the peak
value of the dynamical susceptibility, i.e.,
$\xi_{4}=[\chi_4^{\text{max}}(a,t)]^{1/3}$.   Taking the maximum of
the curves in Fig.\,\ref{fig:ksi28_time_omega_0.05}, the variation
of the correlation length $\xi_{4}$ as a function of the strain
amplitude is presented in the inset of
Fig.\,\ref{fig:ksi28_time_omega_0.05}.   Interestingly, the
correlation length exhibits a distinct maximum at the critical
strain amplitude $\gamma_{0}=0.06$.    These results reveal that the
transition from slow to fast relaxation dynamics reported in
Figs.\,\ref{fig:msd_time_omega}, \ref{fig:norm_modes}, and
\ref{fig:q28_time_omega_0.05} is accompanied by the largest number
of dynamically correlated monomers.

We emphasize that a qualitatively similar behavior of the dynamic
correlation length was observed in the previous study on cyclic
deformation of a binary Lennard-Jones glass~\cite{Priezjev13}.   In
that study, however, the maximum of the dynamical susceptibility
$\chi_4(a,t)$ was computed at the same value of the parameter
$a=0.12\,\sigma$ for all strain amplitudes, and the estimate of the
dynamic correlation length was reported up to a strain amplitude at
which the number of particles involved in a correlated motion is
maximum~\cite{Priezjev13}.    In general, the initial growth of the
dynamic correlation length with increasing strain amplitude during
oscillatory deformation is in marked contrast to the situation in
glassy materials under steady shear, where the relaxation dynamics
becomes more homogeneous with increasing shear
rate~\cite{Tsamados10,Yamamoto12}.    On the other hand, our results
are consistent with the conclusions of the previous study by
Riggleman\,\textit{et al.}~\cite{dePablo10}, who found that at a
constant strain rate deformation of polymer glasses, the relaxation
dynamics is strongly heterogeneous below the yield strain, and after
the onset of flow, the dynamics becomes significantly more
homogeneous.

We now turn to a discussion of the monomer hopping dynamics and the
formation of transient clusters of mobile monomers.    Under cyclic
loading, the motion of a monomer involves thermal vibration within
the cage formed by its neighbors and rapid hopping from one cage to
another.    The cage jumps can be identified by a numerical
algorithm that was originally introduced by Candelier\,\textit{et
al.}~\cite{BiroliPRL09}.    This method, called the
\textit{iterative barycenters separation}, computes the effective
distance between two consecutive segments of a monomer trajectory.
If this distance is larger than the typical cage size then the
trajectory is divided into two subsets.   Using the iterative
procedure, the trajectory of each monomer can be decomposed into
consecutive segments where the displacement of a monomer is
localized within cages formed by its neighbors~\cite{BiroliPRL09}.
This algorithm was used to locate cage jumps in two-dimensional
granular systems~\cite{BiroliPRL09,BiroliEPL10} and supercooled
liquids~\cite{BiroliPRL10}.     More recently, the cage
decomposition algorithm was implemented to identify cage jumps
`on-the-fly' during the simulation run, thus eliminating the need to
store multiple particle configurations~\cite{Rottler13}.

In this work, the monomer trajectories were stored and analyzed
following the cage decomposition method proposed by
Candelier\,\textit{et al.}~\cite{BiroliPRL09}.    Similar to the
implementation of the algorithm used in our previous
study~\cite{Priezjev13}, we first take a part of the monomer
trajectory, cut it in two adjacent segments, and then compute the
effective distance between them.   If this distance is less than the
cage size, $r_c=0.1\,\sigma$, for any two adjacent segments within
the sub-trajectory, then we conclude that the monomer remained
within the cage.   Using this brute-force procedure, we examined all
time intervals $10\,T\leqslant \bigtriangleup t_a \leqslant 100\,T$
for all monomer trajectories.  We found that cage jumps typically
occur during several cycles, and they can be either reversible, when
a monomer jumps back to its previous cage, or irreversible
otherwise.

\begin{figure}[t]
\includegraphics[width=12.cm,angle=0]{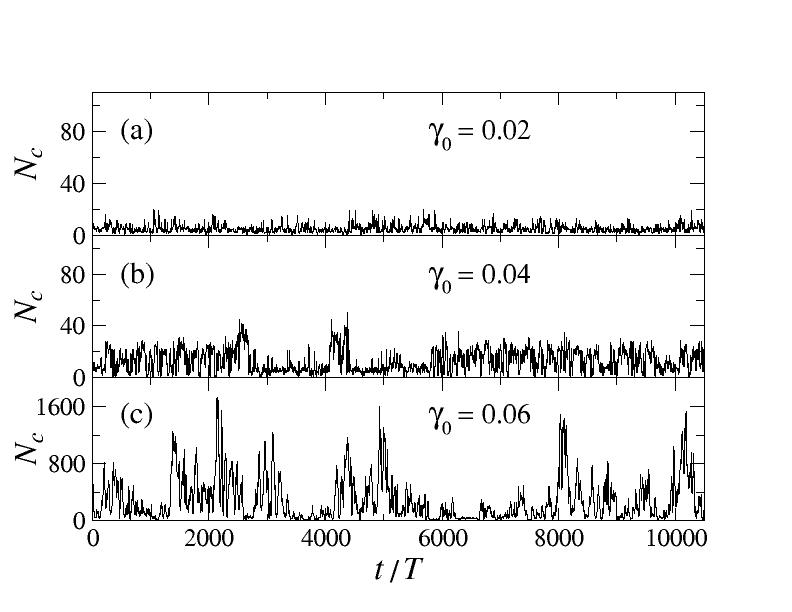}
\caption{(Color online)     The total number of mobile monomers
during cyclic deformation with frequency $\omega\tau=0.05$, period
$T=2\pi/\omega=125.66\,\tau$, and strain amplitudes (a)
$\gamma_{0}=0.02$, (b) $\gamma_{0}=0.04$, and (c) $\gamma_{0}=0.06$.
} \label{fig:time_clusters}
\end{figure}
\begin{figure}[t]
\includegraphics[width=12.cm,angle=0]{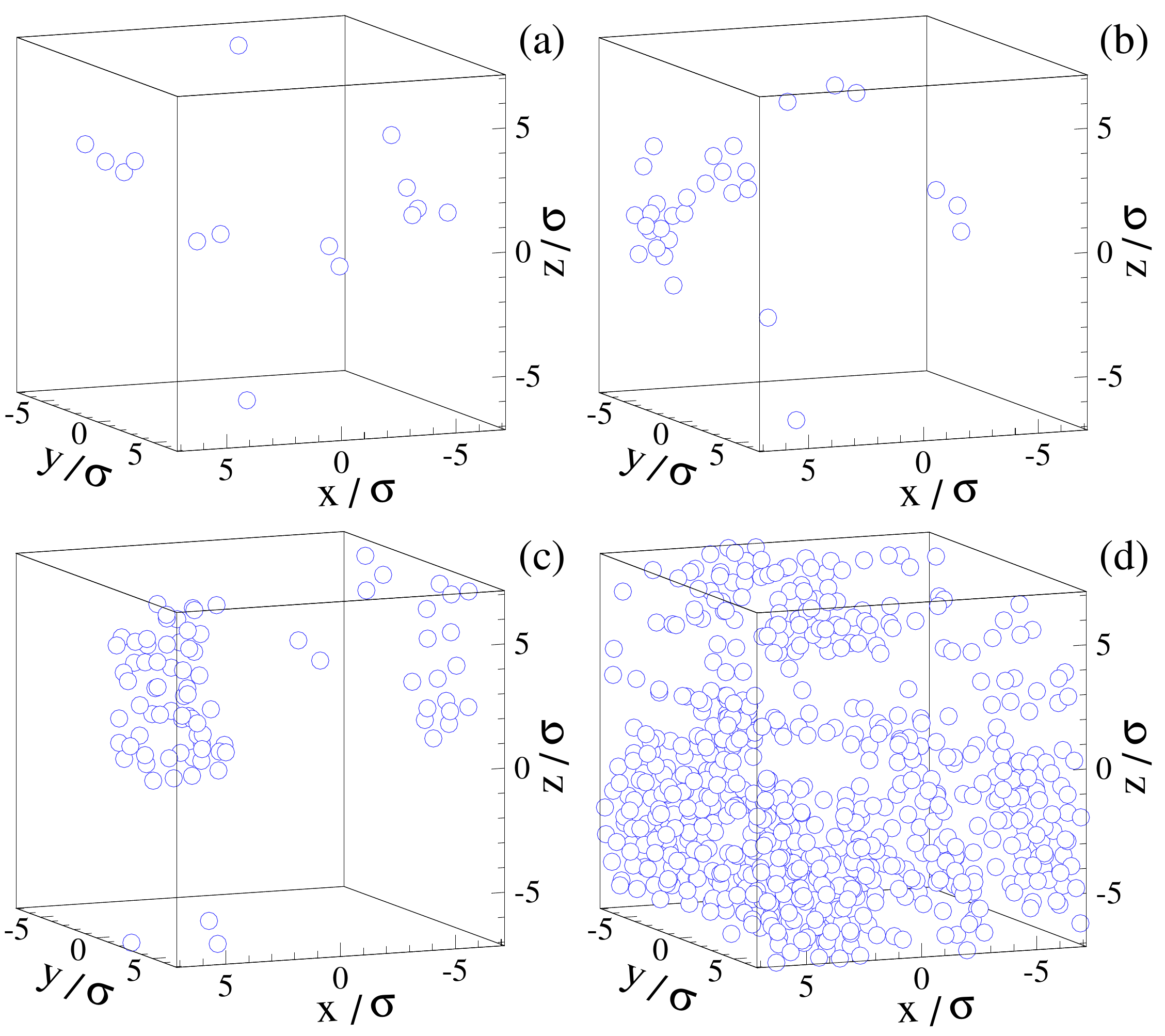}
\caption{(Color online)  Snapshots of mobile monomer configurations
at strain amplitudes (a) $\gamma_{0}=0.03$, (b) $\gamma_{0}=0.04$,
(c) $\gamma_{0}=0.05$, and (d) $\gamma_{0}=0.06$. }
\label{fig:snapshot_clusters}
\end{figure}

The total number of mobile monomers is plotted as a function of time
in Fig.\,\ref{fig:time_clusters} for the strain amplitudes
$\gamma_{0}=0.02,~0.04,~\text{and}~0.06$.   As is evident, in each
case the relaxation dynamics is characterized by sudden bursts in
mobility separated by periods of quiescence.   Note that at the
strain amplitude $\gamma_{0}=0.02$, mobile monomers mostly undergo
reversible jumps without any net displacement (see
Fig.\,\ref{fig:msd_time_omega}), while at $\gamma_{0}=0.06$, the
amplitude of bursts is about half of the total number of monomers in
the system.    Moreover, a visual inspection of snapshots indicates
that mobile monomers tend to form clusters.   Examples of
instantaneous monomer positions during intermittent bursts are shown
in Fig.\,\ref{fig:snapshot_clusters} for different strain
amplitudes.    It can be seen that the clusters have a relatively
compact structure, although several monomers appear to be isolated.
Notice the formation of a large cluster at $\gamma_{0}=0.06$ that
spans the whole system.   As the cyclic deformation continues, the
number of mobile monomers in a large cluster decreases typically to
a few monomers that undergo reversible jumps until the emergence of
the next cluster.   Thus, the spatio-temporal clusterization
algorithm~\cite{BiroliPRL09} would identify only several large-size
clusters, rendering their statistics unreliable.

\begin{figure}[t]
\includegraphics[width=12.cm,angle=0]{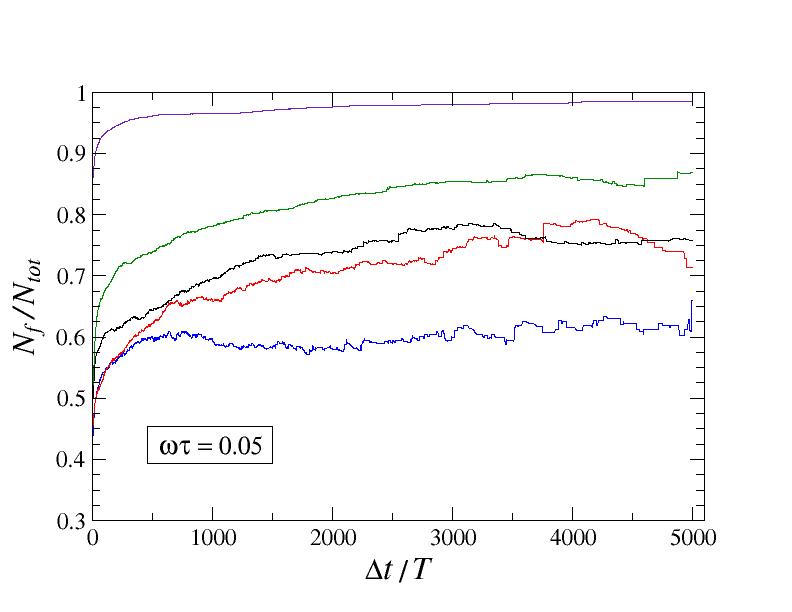}
\caption{(Color online)   The ratio of dynamically facilitated
mobile monomers and the total number of mobile monomers as a
function of the time interval preceding cage jumps.  The oscillation
frequency is $\omega\tau=0.05$ and strain amplitudes are
$\gamma_{0}=0.02,~0.03,~0.04,~0.05,~\text{and}~0.06$ (from bottom to
top).  } \label{fig:dyn_facil}
\end{figure}

A number of previous studies have explored the effect of dynamic
facilitation in glassy materials and concluded that a particle has a
higher probability to become mobile if it has a neighboring particle
that was previously
mobile~\cite{Chandler03,Glotzer05,BiroliEPL10,Priezjev13}.  Here, we
analyzed a trajectory of each monomer, using the cage decomposition
algorithm described above, and identified cage jumps and time
intervals when a monomer remained within the cage.    If a monomer
escaped its cage and had at least one nearest-neighbor that was
mobile sometime during the time interval $\Delta\,t$ preceding the
cage jump, then this hop event was considered to be facilitated by
the neighbors.  The ratio of dynamically facilitated mobile monomers
and the total number of mobile monomers is plotted in
Fig.\,\ref{fig:dyn_facil} as a function of the time interval
preceding cage jumps.   It can be observed that the ratio
$N_{f}/N_{tot}$ increases rapidly and appears to saturate after
about $2000$ cycles.   With increasing strain amplitude, the
fraction of dynamically facilitated mobile monomers increases.    It
is perhaps not surprising that the ratio is nearly one for
$\gamma_{0}=0.06$ because most of the monomers undergo cage jumps
during the time interval $10^4\,T$ and thus influence hopping of
their neighbors.    The fact that the ratio is about $0.6$ for the
strain amplitude $\gamma_{0}=0.02$, at which the self-correlation
function remains nearly constant (see
Fig.\,\ref{fig:q28_time_omega_0.05}), suggests that reversible cage
jumps are either spatially isolated or clustered in small groups at
the same locations during the reported time interval.

\section{Conclusions}

In summary, molecular dynamics simulations were carried out to
investigate structural relaxation and dynamical heterogeneity in a
model polymer glass under oscillatory shear strain.    We used a
standard bead-spring representation of linear polymer chains below
the entanglement regime.    To probe the microscopic relaxation
dynamics, we examined the mean-square displacement of monomers, the
autocorrelation function of normal modes, as well as the
self-overlap order parameter and dynamical susceptibility.

It was found that the segmental mobility is unaffected by the
time-periodic deformation at small strain amplitudes, whereas the
relaxation time of polymer chains becomes less than about a hundred
oscillation periods at strain amplitudes above a few percent.    By
computing the peak value of the dynamical susceptibility, we
estimated the dynamical correlation length that was found to exhibit
a distinct maximum at the critical strain amplitude.    Therefore,
it was concluded that the transition from slow to fast relaxation
dynamics is associated with the largest number of monomers involved
in the correlated motion.

The post-processing analysis of all monomer trajectories, based on
the cage decomposition algorithm~\cite{BiroliPRL09}, indicated that
mobile monomers tend to aggregate into transient clusters.     It
was observed that the typical cluster size during intermittent
bursts increases at larger strain amplitudes, which is in agreement
with findings of the previous study on cyclic loading of a binary
Lennard-Jones glass~\cite{Priezjev13}.    The effect of dynamic
facilitation of mobile monomers by their neighbors becomes more
pronounced with increasing strain amplitude.

In the future, it would be instructive to perform a finite-size
scaling analysis of the dynamic correlation length in the vicinity
of the critical strain amplitude and to explore the influence of
oscillation frequency on the structural relaxation dynamics in
polymer glasses.

\section*{Acknowledgments}

Financial support from the National Science Foundation (Grant No.
CBET-1033662) is gratefully acknowledged.  Computational work in
support of this research was performed at Michigan State
University's High Performance Computing Facility and the Ohio
Supercomputer Center.



\bibliographystyle{prsty}

\begin{thebibliography}{99}


\bibitem{Meijer05}    H.~E.~H. Meijer and L.~E. Govaert,
                      Prog. Polym. Sci. {\bf 30}, 915 (2005).

\bibitem{Struik78}    L. C. E. Struik,
                      {\it Physical Aging in Amorphous Polymers and Other Materials}
                      (Elsevier, Amsterdam, 1978).

\bibitem{Eyring36}    H. Eyring, J. Chem. Phys. {\bf 4}, 283 (1936).

\bibitem{BarratCh8}   J.-L. Barrat and A. Lemaitre,
                      ``Heterogeneities in amorphous systems under shear," in
                      \textit{Dynamical Heterogeneities in Glasses, Colloids, and Granular Media}
                      (Oxford University Press, 2011), chap. 8.

\bibitem{Hoy06}       R.~S. Hoy and M.~O. Robbins,
                      J. Polymer Science Part B: Polymer Physics {\bf 44}, 3487 (2006).

\bibitem{GlotNat99}   C. Bennemann, C. Donati, J. Baschnagel, and S.~C. Glotzer,
                      Nature {\bf 399}, 246 (1999).

\bibitem{GlotStar01}  Y. Gebremichael, T.~B. Schroder, F.~W. Starr, S.~C. Glotzer,
                      Phys. Rev. E {\bf 64}, 051503 (2001).

\bibitem{Aichele03}   M. Aichele, Y. Gebremichael, F.~W. Starr, J. Baschnagel, S.~C. Glotzer,
                      J. Chem. Phys. {\bf 119}, 5290 (2003).

\bibitem{Rottler13}   A. Smessaert and J. Rottler,
                      Phys. Rev. E {\bf 88}, 022314 (2013).

\bibitem{Gleason00}   L.~S. Loo, R.~E. Cohen, K.~K. Gleason,
                      Science {\bf 288}, 116 (2000).

\bibitem{Ediger08}    H.-N. Lee, K. Paeng, S.~F. Swallen, M.~D. Ediger,
                      J. Chem. Phys. {\bf 128}, 134902 (2008).

\bibitem{Ediger09}    H.-N. Lee, K. Paeng, S.~F. Swallen, M.~D. Ediger,
                      Science {\bf 323}, 231 (2009).

\bibitem{Rutledge02}  F.~M. Capaldi and M.~C. Boyce, and G.~C. Rutledge,
                      Phys. Rev. Lett. {\bf 89}, 175505 (2002).

\bibitem{Lyulin05}    A.~V. Lyulin, B. Vorselaars, M.~A. Mazo, N.~K. Balabaev, M.~A.~J. Michels,
                      Europhys. Lett. {\bf 71}, 618 (2005).

\bibitem{Papa08}      G.~J. Papakonstantopoulos, R.~A. Riggleman, J.-L. Barrat, J.~J. de Pablo,
                      Phys. Rev. E {\bf 77}, 041502 (2008).

\bibitem{dePablo10}   R.~A. Riggleman, H.-N. Lee, M.~D. Ediger, and J.~J. de Pablo,
                      Soft Matter {\bf 6}, 287 (2010).

\bibitem{RottlerPRL10} M. Warren and J. Rottler,
                      Phys. Rev. Lett. {\bf 104}, 205501 (2010).

\bibitem{RottlerJCP10} M. Warren and J. Rottler,
                      J. Chem. Phys. {\bf 133}, 164513 (2010).
\bibitem{BiroliPRL09} R. Candelier, O. Dauchot, and G. Biroli,
                      Phys. Rev. Lett. {\bf 102}, 088001 (2009).

\bibitem{Sastry13}    D. Fiocco, G. Foffi, S. Sastry,
                      Phys. Rev. E {\bf 88}, 020301(R) (2013).

\bibitem{Priezjev13}  N.~V. Priezjev,
                      Phys. Rev. E {\bf 87}, 052302 (2013).

\bibitem{OHern13}     C.~F. Schreck, R.~S. Hoy, M.~D. Shattuck, and C.~S. O'Hern,
                      arXiv:1301.7492 (2013).

\bibitem{Lyulin13}    D. Hudzinskyy, M.~A.~J. Michels, A.~V. Lyulin,
                      Macromol. Theory Simul. {\bf 22}, 71 (2013).

\bibitem{Reichhardt13} I. Regev, T. Lookman, and C. Reichhardt, arXiv:1301.7479 (2013).

\bibitem{Arratia13}   N.~C. Keim and P.~E. Arratia, arXiv:1308.6806 (2013).

\bibitem{Kremer90}    K. Kremer and G.~S. Grest,
                      J. Chem. Phys. {\bf 92}, 5057 (1990).

\bibitem{Bird87}      R.~B. Bird, C.~F. Curtiss, R.~C. Armstrong, and O. Hassager,
                      {\it Dynamics of Polymeric Liquids}, 2nd ed. (Wiley, New York, 1987).

\bibitem{Binder98}    C. Bennemann, W. Paul, K. Binder, and B. Dunweg,
                      Phys. Rev. E {\bf 57}, 843 (1998).

\bibitem{Evans92}     D.~J. Evans and G.~P. Morriss,
                      {\it Statistical Mechanics of Nonequilibrium Liquids} (Academic Press, London, 1990).

\bibitem{Allen87}     M.~P. Allen and D.~J. Tildesley,
                      {\it Computer Simulation of Liquids} (Clarendon, Oxford, 1987).

\bibitem{Paul04}      W. Paul and G.~D. Smith,
                      Rep. Prog. Phys. {\bf 67}, 1117 (2004).

\bibitem{Verdier66}   P.~H. Verdier,
                      J. Chem. Phys. {\bf 45}, 2118 (1966).

\bibitem{Varnik05}    J. Baschnagel and F. Varnik,
                      J. Phys.: Condens. Matter {\bf 17}, R851 (2005).

\bibitem{Berthier11}  L. Berthier and G. Biroli,
                      Rev. Mod. Phys. {\bf 83}, 587 (2011).

\bibitem{BiroliPRL05} O. Dauchot, G. Marty, and G. Biroli,
                      Phys. Rev. Lett. {\bf 95}, 265701 (2005).

\bibitem{Glotzer00}   S.~C. Glotzer, V.~N. Novikov, and T.~B. Schroder,
                      J. Chem. Phys. {\bf 112}, 509 (2000).

\bibitem{Tsamados10}  M. Tsamados,
                      Eur. Phys. J. E {\bf 32}, 165 (2010).

\bibitem{Yamamoto12}  H. Mizuno and R. Yamamoto,
                      J. Chem. Phys. {\bf 136}, 084505 (2012).

\bibitem{BiroliEPL10} R. Candelier, O. Dauchot, and G. Biroli,
                      Europhys. Lett. {\bf 92}, 24003 (2010).

\bibitem{BiroliPRL10} R. Candelier, A. Widmer-Cooper, J.~K. Kummerfeld,
                      O. Dauchot, G. Biroli, P. Harrowell, and D.~R. Reichman,
                      Phys. Rev. Lett. {\bf 105}, 135702 (2010).

\bibitem{Chandler03}  J.~P. Garrahan and D. Chandler,
                      Proc. Natl. Acad. Sci. USA {\bf 100}, 9710 (2003).

\bibitem{Glotzer05}   M.~N.~J. Bergroth, M. Vogel, and S.~C. Glotzer,
                      J. Phys. Chem. B {\bf 109}, 6748 (2005).

\end{thebibliography}

\end{document}